\documentclass[twocolumn,pra,superscriptaddress]{revtex4}

\usepackage{amsmath}
\usepackage{amssymb}

\newcommand{\rme}{{\rm e}}
\newcommand{\rmi}{{\rm i}}

\newcommand{\eps}{\varepsilon}
\newcommand{\la}{\langle}

\newcommand{\ra}{\rangle}

\newcommand{\Idone}{\boldsymbol{1}}

\begin{document}

\title{Decoherence alias Loschmidt echo of the environment}

\author{T. Gorin}
\affiliation{Theoretische Quantendynamik, Albert-Ludwigs-Universit\" at, 
   Hermann-Herder-Str. 3, 79104 Freiburg, Germany}
\author{T. Prosen}
\affiliation{Physics Department, Faculty of Mathematics and Physics,
   University of Ljubljana, Jadranska 19, SI-1000 Ljubljana, Slovenia}
\author{T. H. Seligman}
\affiliation{Centro Internacional de Ciencias, C.P. 62131 Cuernavaca,
   Morelos Mexico}
\affiliation{Centro de Ciencias F\'\i sicas, University of Mexico (UNAM),
   C.P. 62210 Cuernavaca, Morelos, Mexico}
\author{W. T. Strunz}
\affiliation{Theoretische Quantendynamik, Albert-Ludwigs-Universit\" at,
   Hermann-Herder-Str. 3, 79104 Freiburg, Germany}


\begin{abstract}
Entanglement between a quantum system and its environment leads to
loss of coherence in the former. In general, the temporal fate
of coherences is complicated.
Here, we establish the connection between decoherence of a central system and
fidelity decay in the environment for a variety of situations,
including both, energy conserving and dissipative couplings.
We show how properties of unitary time evolution of the environment can be
inferred from the non-unitary evolution of coherences in the central system.
This opens up promising ways for measuring Loschmidt echoes in a variety of
situations.
\end{abstract}

\maketitle

\section{Introduction}

Quantum Loschmidt echoes~\cite{Pasta2000}
have received a large amount of theoretical and experimental attention
in recent years~\cite{theory,Prosen:rev,GPS04,experiment}.  Such echoes
are obtained by propagating an initial state for some time $t$ forwards and
then backwards in time. In the ideal situation where the forward and backward
evolutions are the same, the system ends up in its initial state at time $2t$.
However, in reality the forward and backward evolution are distorted by
inherently uncontrollable perturbations. These deviations typically add up in
the course of the evolution, which results in a final state being notably
different from the initial state. A natural measure for this difference is
the overlap of both states, the {\it fidelity amplitude}. Its absolute value
squared is the {\it fidelity}. Experimentally, this concept of Loschmidt
echoes has been widely used in connection with nuclear magnetic resonance
spectroscopy, photon echoes and wave packet echoes of trapped
atoms~\cite{experiment}.

Alternatively, fidelity can be looked at from a different viewpoint. Namely,
one may consider two identical initial states being propagated according to
slightly different Hamiltonians. Then, after time $t$ the overlap between the
two states will no longer be equal to one. Formally this quantity is
again a fidelity amplitude. 
In this picture the relation between the separation of nearby trajectories in
classical dynamics (as a measure of chaos) and fidelity decrease becomes most
transparent~\cite{theory,Prosen:rev,GPS04,experiment}.

While fidelity is based on the unitary time evolution of the quantum system
of interest, {\em decoherence} arises due to the coupling to additional
``environmental'' degrees of freedom, {\em i.e.} due to the growing
entanglement between that system and its ``environment''.
Decoherence as a dynamical phenomenon has received growing attention
in the last few years~\cite{Zeh, Zurek}. The reason is obvious:
For newly emerging quantum technologies, such as
quantum cryptography and quantum computing, or quantum information
processing in general, the stability of quantum coherence is
fundamental~\cite{Nielsen}.
Decoherence is {\it the} obstacle that has
to be overcome for these technologies to prove successful.
This requires 
a clear understanding of mechanisms and time scales involved.

In this work we investigate situations where decoherence in the central system
can be related to fidelity decay in the environment. We shall show that this
connection between apparently unrelated research areas is quite general.
The principal idea, 
is to use an internal degree of freedom both to create the
difference between the two Hamiltonians involved and
to monitor the fidelity decay in the course of the evolution.
An experimental configuration which allows to realize a fidelity
measurement of  this type has been proposed in~\cite{GCZ97,GCZ97b}.
We investigate various situations where it is possible to interpret
coherences (off-diagonal elements of the reduced density operator) in
one subsystem as fidelity amplitudes of unitary, perturbed dynamics in
the other. The strength of the perturbation may be related to the
``distance'' of the initially superposed states, as will be explained below.

The argument is based on the unitary evolution in the
product Hilbert space ${\cal H} = {\cal H}^{\rm c}\otimes {\cal H}^{\rm e}$,
of the Hilbert spaces  for the central system (c) and the environment (e),
respectively. We consider a total Hamiltonian of the form
\begin{equation}\label{generalH}
H = H^{\rm c}+ H^{\rm e} + H^{\rm int}
\end{equation}
consisting of two Hamiltonians that describe the two subsystems separately,
and an interaction term $H^{\rm int}$,
for which we shall consider different forms, as
specified below. Note that up to this point, the designations as
``environment'' and ``central system'' are purely conventional. The only
important point is the existence of two spaces.
Over one of these, {\it i.e.} the ``environment'',  we shall execute
partial traces to consider the entanglement between the two spaces in terms
of the off-diagonal matrix elements of the density matrix in the other space,
{\it i.e.} the ``central system''.

 In section~\ref{E} we consider a coupling $H^{\rm int}$ that conserves
the energy of the central system. The environmental influence
is thus not of the dissipative type; still, phase relations in the
central system will be disturbed and thus coherences lost.
This setting is a generalization of recent 
proposals and experimental realizations in the literature.
   In section~\ref{A} we turn our attention to the damped harmonic oscillator,
{\em i.e.} to the so-called amplitude coupling between a central oscillator
and a ``bath'' of environmental oscillators. We show that a similar
connection between decoherence and fidelity decay may hold even
in this dissipative case, where the coupling ($H^{\rm int}$) and the
Hamiltonian of the central system ($H^{\rm c}$) do not commute. The
famous Paris decoherence experiment~\cite{Bru96} may thus be
interpreted as an environmental ``Loschmidt-echo''-experiment.
   Finally, in section~\ref{S} we consider more general situations where the
relation between decoherence and environmental echo is
only approximately valid. 
This is the case, if 
decoherence alias fidelity decay is fast, compared to typical time scales in the 
isolated central system.

\section{\label{E} Energy conserving coupling -- ``dephasing''}

An energy conserving coupling for the central system
is realized when the coupling term in (\ref{generalH}) is of the form:
\begin{equation}
H^{\rm int} = \sum_j |\phi_j\ra\; \la\phi_j| \otimes V^{\rm e}_j \; ,
\label{V-1}\end{equation}
where the $\{\phi_j\}$ form a complete set of eigenstates of $H^{\rm c}$ (for
convenience, we shall assume that the spectrum of $H^{\rm c}$ is discrete).
In the eigenbasis representation:
$H^{\rm c}= \sum_j |\phi_j\ra\,\eps_j\, \la\phi_j|$,
the full Hamiltonian may be written as:
\begin{eqnarray}
H &=& \sum_j |\phi_j\ra\, \eps_j\, \la\phi_j| \otimes \Idone +
   \Idone \otimes H^{\rm e} + \sum_j |\phi_j\ra\, \la\phi_j|
   \otimes V^{\rm e}_j \nonumber\\
&=& \sum_j |\phi_j\ra\, \la\phi_j| \otimes \left[ \eps_j\; \Idone +
       H^{\rm e} + V^{\rm e}_j \right] \; .
\label{E:Ham}\end{eqnarray}
As the Hamiltonian $H^{\rm c}$ commutes with $H$,
the energy of the central system is conserved. Hence, the eigenstates of
$H$ must be separable: $|\Psi\ra = |\phi_j\ra \otimes |\chi_{j}^\alpha\ra$,
where the wave functions $|\chi_{j}^\alpha\ra$ satisfy the $j$-dependent
Schr\" odinger equation
\begin{equation}
\left[ H^{\rm e} + V^{\rm e}_j + \eps_j \right] |\chi_{j}^\alpha \ra
   = E_j^\alpha \; |\chi_{j}^\alpha\ra
\end{equation}
in the Hilbert space of the environment.

\subsubsection*{Time evolution and fidelity}

Since $H^{\rm c}$ is a constant of motion, an initial product state
$|\Psi_0\ra = |\phi_j\ra\otimes|\chi_0\ra$ with an eigenfunction $|\phi_j\ra$
of $H^{\rm c}$ will remain a product state for all times. We find
\begin{equation}
|\Psi(t)\ra = \rme^{-\rmi\, \eps_j\, t}\; |\phi_j\ra \otimes |\chi_j(t)\ra
\label{E:Psi_product}\end{equation}
with the environmental state $|\chi_j(t)\ra$ obeying the $j$-dependent
Schr\"odinger equation
\begin{equation}
\rmi\;\hbar\,\partial_t\; |\chi_j(t)\ra = \left[ H^{\rm e} + V^{\rm e}_j
   \right] |\chi_j(t)\ra
\label{jSchreqation}\end{equation}
in the Hilbert space of the environment.
Clearly, the initial state $|\chi_j(0)\ra =  |\chi_0\ra$ is independent of $j$.

In general, an initially separable state
$|\Psi_0\ra = |\phi_0\ra\otimes|\chi_0\ra$ will not remain
separable under time evolution. Using the eigenbasis of $H^{\rm c}$ we write
$|\phi_0\ra = \sum_j a_j |\phi_j\ra$ and find from the previous
considerations the entangled state
\begin{equation}
|\Psi(t)\ra = \sum_j a_j\; \rme^{-\rmi\, \eps_j\, t}\; |\phi_j\ra
\otimes |\chi_j(t)\ra.
\label{E:Psi}\end{equation}

Crucially, the ``perturbing potential'' $V^{\rm e}_j$ that governs the evolution
of the  environmental states depends on
the choice of the $H^{\rm c}$-eigenstate $|\phi_j\ra$.
We also see that $|\chi_j(t)\ra$
evolves unitarily, so that its norm is conserved. From equation~(\ref{E:Psi})
we compute the reduced density matrix $\varrho^{\rm c}(t)$ in the central
system:
\begin{eqnarray}
 \varrho^{\rm c}(t) &=& {\rm Tr}_{\rm e}\; |\Psi(t)\ra\, \la\Psi(t)|
 =  {\rm Tr}_{\rm e}\sum_{jk} a_j\, a_k^*\, \rme^{-\rmi\, (\eps_j-\eps_k)\, t}
\nonumber\\
&&\qquad\times |\phi_j\ra\, \la\phi_k| \otimes |\chi_j(t)\ra\, \la\chi_k(t)| \; .
\end{eqnarray}
Hence, coherences between eigenstates $|\phi_j\ra,|\phi_k\ra$ of the
central system are given by the matrix elements:
\begin{equation}\label{rhojk}
\varrho^{\rm c}_{jk}(t) =
\rme^{-\rmi\, (\eps_j-\eps_k)\, t}\;
   \la\chi_k(t)|\chi_j(t)\ra\;
\varrho^{\rm c}_{jk}(0)
 \; .
\end{equation}

The decay of coherences is thus determined by the decay of
the fidelity amplitude in the Hilbert space of
the environment, for which we can write
\begin{equation}
\la\chi_k(t)|\chi_j(t)\ra = \la\chi_0|M(t)|\chi_0\ra \; ,
\label{echo-1}\end{equation}
where $M(t)= \tilde U_0(-t)\; \tilde U(t)$ is a so called echo 
operator~\cite{Prosen:rev,Prosen_etal}, while $\tilde U_0(t)$
and $\tilde U(t)$ are the respective evolution operators for the Hamiltonians:
\begin{equation}\label{echoHamiltonian}
\tilde H_0 = H^{\rm e} + V^{\rm e}_k \qquad
\tilde H = H_0 + V^{\rm e}_j - V^{\rm e}_k \; .
\end{equation}
Note that each non-diagonal matrix element of $\varrho^{\rm c}$ involves a
different echo-operator with (slightly) different $\tilde H_0$ and
$\tilde H$. However, in many cases the initial coefficients $a_j$ as well as
the phases $\exp[-\rmi\, (\eps_j-\eps_k)\, t]$ can be controlled quite
precisely (see section~\ref{X}), so that coherences and the corresponding
fidelities are readily identified in actual experiments. \\

We may finally mention a special situation of interest,
where the environment factor of the separable interaction is simply
proportional to the
Hamiltonian of the environment, {\it i.e.}
$V^{\rm e}_j = f_j\, H^{\rm e}$ with some real number $f_j$.
In this case equation~(\ref{echo-1}) simplifies to
\begin{equation}
\la\chi_k(t)|\chi_j(t)\ra =\la\chi_0|  e^{-i t
[f_k- f_j]\, H^{\rm e}}\, |\chi_0\ra \; ,
\end{equation}
which is the autocorrelation function in the environment of
$\chi_0$ under a $k$ and $j$-dependent rescaled time evolution.

\subsection{\label{X} Experimental realizations with trapped atoms}

Experimental setups which allow to realize such a scheme have been
proposed in ~\cite{GCZ97,GCZ97b}, based on
a single cold ion in a trapping potential involving two different
electronic states $|1\ra$ and $|2\ra$. The electronic
states play the role of the ``central system'', while
-- in our terminology -- the
center-of-mass motion of the ion should be identified with the
``environmental'' degrees of freedom. Here, the dynamics of interest is
the motion of the ion in the trap (eventually, one may wish to find out
whether it corresponds to classically chaotic or integrable motion).
The proposal is based on
an initial state involving a coherent superposition of
both internal states,
\begin{equation}
\Psi(0)= 2^{-1/2}\; \big(\, |1\ra + |2\ra \big)\; \otimes |\chi_0\ra \; .
\label{E:Psi0}\end{equation}
Here, $|\chi_0\ra$ being the initial motional state of the ion,
for instance a coherent state.
The ion evolves in the trap potential for some time $t$ under
the influence of an internal state-dependent potential, as explained
previously. Physically, this is achieved with the help of a constant or pulsed
off-resonant laser field (ac stark effect).
After some time $t$, the coherence $\varrho^{\rm c}_{12}(t)$ may
be measured using Ramsey techniques~\cite{Ramsey_techniques}. A recent
experiment~\cite{Arcy03} with laser cooled Cs atoms exposed to the
gravitational field and pulses of a standing wave of off-resonant light
is close to a realization of such concepts. The authors use
two hyperfine levels as internal ``central'' system, and
propose to measure fidelity decay in a chaotic system. \\

Finally, let us
mention an experiment where a quantity closely related
to an echo fidelity is measured through the loss of coherence
in a ``central system''. In~\cite{AndDav03} the authors
investigate ultra cold $^{85}$Rb atoms
in an optical dipole trap. Using our terminology, the ``central system''
consists of internal electronic levels, while
the center-of-mass motion of the atoms plays the role of the
``environment''. 
Starting with the initial state of equation~(\ref{E:Psi0}), and applying
an additional $\pi$-pulse right in the middle of the time evolution,
one obtains for the coherences in the ``central system'':
\begin{equation}
\varrho^{\rm c}_{12}(t) =
\la\chi_0|
     U_2^\dagger\, U_1^\dagger\, U_2\, U_1 |\chi_0\ra
\;\varrho^{\rm c}_{12}(0) \; .
\end{equation}
Here, the echo operator is replaced by a product of four evolution
operators over half the time interval, $t/2$, while the phases
originating from the evolution of the central system have
canceled.
This particular variant of the echo-operator has the advantage that the
echo-signal is insensitive to the dephasing of different motional
eigenstates of the atoms. In the experiment, it allows to observe an
echo, even though about $10^6$ states are thermally populated. Ultimately,
the decay of the response is related to the detuning of the trap laser with
respect to the different hyperfine states of the atoms.

\section{\label{A} Amplitude coupling between harmonic oscillators}


In this section we consider a particular dissipative system, namely the
famous quantum optical damped harmonic oscillator. Both, central system
and environment consist of harmonic oscillators; the coupling is bilinear
in annihilation and creation operators:
\begin{eqnarray}\label{Hoscis}
H & = & H^{\rm c} + H^{\rm e} + H^{\rm int} \\ \nonumber
& = & \hbar\Omega\; a^{\dagger}a +
    \sum_{\lambda}\hbar\omega_{\lambda}b_{\lambda}^{\dagger}b_{\lambda}+
    \sum\limits_{\lambda}\hbar g_{\lambda}\left( ab_{\lambda}^{\dagger}
     + a^{\dagger}b_{\lambda} \right).
\end{eqnarray}
Remarkably, despite so-called ``amplitude coupling'', this model allows for
the correspondence between fidelity decay and decoherence. Moreover,
the beautiful Paris decoherence experiment
of a microwave field in a superconducting cavity ~\cite{Bru96}
is adequately described by the Hamiltonian (\ref{Hoscis}).
In the light of the results to be shown, this decoherence experiment
(for the central oscillator)
may now also be interpreted as a ``fidelity decay'' experiment
for the environment.
A detailed theoretical description of the experiment
may be found in~\cite{Dav96}.

As in the case of energy conserving coupling considered
previously, we have to identify
product state solutions of the dynamics. For
Hamiltonian (\ref{Hoscis}), they are given by products of coherent states.
It is easy to see that
with $|z\rangle = \exp\{-\frac{1}{2}|z|^2 + za^\dagger\}|0\rangle$ for the
central system and similarly defined coherent states $|\beta_\lambda\rangle$
for the oscillators of the environment, the product state
\begin{eqnarray}
|\Psi(t)\rangle &=& |z(t)\rangle\otimes|\beta_1(t)\rangle
\otimes|\beta_2(t)\rangle\otimes\cdots
\otimes|\beta_\lambda(t)\rangle\otimes\cdots  \nonumber\\
&\equiv & |z(t)\rangle\otimes|B(t)\rangle
\end{eqnarray}
is a solution of the Schr\"odinger equation. This holds true provided the
coherent state labels follow the classical equations of motion:
\begin{eqnarray}\label{claseq}
i\partial_t z(t) & = & \Omega z(t) + \sum_\lambda g_\lambda b_\lambda(t)
\\ \nonumber
i\partial_t \beta_\lambda(t) & = & \omega_\lambda \beta_\lambda(t)
+ g_\lambda z(t).
\end{eqnarray}

Assume, for simplicity (and also in very good agreement with
experiment), a zero temperature environment such that all
$\beta_\lambda(0) = 0$. Formal integration leads
to $\beta_\lambda(t) = -i g_\lambda\int_0^t ds\,e^{-i\omega_\lambda(t-s)}
z(s)$. For the central system we find the effective equation
\begin{equation}\label{zerot}
{\dot z}(t) + i\Omega z(t) + \int_0^t ds\,\alpha(t-s) z(s) = 0
\end{equation}
with the zero temperature bath correlation function
$\alpha(t-s) = \sum_\lambda |g_\lambda|^2e^{-i\omega_\lambda(t-s)}$.
The actual experiment is well described by the Markov approximation
which amounts to the replacement $\alpha(t-s)=\gamma\delta(t-s)$. Then
$z(t) = \exp\{-i\Omega t -\frac{\gamma}{2} t\}$ displays the
expected damped harmonic motion of the central oscillator.
For the following argument, however, no such approximation is
necessary.

We choose to investigate the fate
of an initial ``Schr\"odinger cat state'' of the central oscillator
coupled to the environmental vacuum,
\begin{equation}\label{initial}
     |\Psi(0)\rangle =
     \frac{1}{\sqrt{2}}(|z_1(0)\rangle + |z_2(0)\rangle)\otimes
      |0\rangle.
\end{equation}
Here, for simplicity, we assume $|z_1(0)-z_2(0)|\gg 1$ such that
$\langle z_1(0)|z_2(0)\rangle\approx 0$ which simplifies the
normalization in (\ref{initial}). Linearity demands that
the total state evolves into the entangled state
\begin{equation}\label{entangledPsi}
     |\Psi(t)\rangle
    = \frac{1}{\sqrt{2}}|z_1(t)\rangle\otimes |B_1(t)\rangle
     + \frac{1}{\sqrt{2}}|z_2(t)\rangle\otimes |B_2(t)\rangle \; ,
\end{equation}
where we denote with $|B_i\rangle =
|\beta_1^{(i)}\rangle\otimes|\beta_2^{(i)}\rangle\otimes\cdots
\otimes|\beta_\lambda^{(i)}\rangle\otimes\cdots$ the
environmental state corresponding to the initial state
$|z_i(0)\ra \otimes |0\ra$.
The coherent state labels in (\ref{entangledPsi}) evolve according to the
classical equations (\ref{claseq}) with initial conditions
$\{z_1(0),\beta_\lambda^{(1)}=0\}$ and
$\{z_2(0),\beta_\lambda^{(2)}=0\}$ respectively.
The reduced density operator of the central system
$\varrho^{\rm c} = {\rm Tr}_{\rm e}\, |\Psi\rangle\langle\Psi|$
is easily determined from the total state (\ref{entangledPsi}) and
we find
\begin{eqnarray}\label{reddens}\nonumber
\varrho^{\rm c}(t) & = &  \frac{1}{2}|z_1(t)\rangle \langle  z_1(t)|
 + \frac{1}{2}|z_2(t)\rangle \langle z_2(t)| \\ \nonumber
 &&\qquad + \frac{1}{2} \langle B_2(t)|B_1(t)
          \rangle |z_1(t)\rangle \langle z_2(t)| \\ 
&&\qquad + \frac{1}{2}\langle B_1(t)|B_2(t) \rangle |z_2(t)\rangle \langle
 z_1(t)|\; .
\end{eqnarray}
Clearly, the time dependence of the coherence between the superposed states
is determined by the fidelity $\langle B_2(t)|B_1(t) \rangle$ of the
corresponding environmental states.

As in the case of the energy-conserving coupling,
the decaying fidelity may be interpreted as an echo fidelity.
To see that, first notice
that apart from an irrelevant phase
with ${\dot\phi}_j(t) = -\frac{1}{2}\sum_\lambda g_\lambda
(z_j(t)b_\lambda^{(j)}(t) + c.c.)$, the environmental states
$|B_j(t)\rangle$ satisfy Schr\"odinger's equation with
time dependent environment Hamiltonian
\begin{equation}\label{envhamil}
H^{\rm e}_j =
    \sum_{\lambda}\hbar\omega_{\lambda}b_{\lambda}^{\dagger}b_{\lambda}+
    \sum\limits_{\lambda}\hbar g_{\lambda}\left( z_j(t) b_{\lambda}^{\dagger}
     + z_j^*(t)b_{\lambda} \right),
\end{equation}
describing harmonic oscillations ``driven'' by the amplitude
$z_j(t)$ of the damped central oscillator
as determined from (\ref{zerot}). Its initial value $z_j(0)$ arises
from the initial state of the central system (\ref{initial}). Different
initial coherent states $|z_j(0)\rangle$
lead to different Hamiltonians $H^{\rm e}_j$ in (\ref{envhamil})
and thus
give different time evolutions of the environmental states.
Similar to the equations (\ref{rhojk}), (\ref{echo-1}),
and (\ref{echoHamiltonian}) we may write
\begin{eqnarray}
\varrho_{12}^{\rm c}(t) &=& \langle B_2(t)|B_1(t) \rangle\;
\varrho_{12}^{\rm c}(0) \nonumber\\
&=& e^{-\rmi(\phi_1(t)-\phi_2(t))}\langle 0 |{\tilde U}_0^\dagger(t)
{\tilde U}(t) |0\rangle\; \varrho_{12}^{\rm c}(0)
\end{eqnarray}
with the propagators arising from the Hamiltonians
\begin{eqnarray}\label{echoH}
{\tilde H}_0 &=& H^{\rm e}_1 \nonumber\\
{\tilde H} &=& {\tilde H}_0 +\left\{(z_2(t)-z_1(t))
    \sum\limits_{\lambda}\hbar g_{\lambda}
b_{\lambda}^{\dagger} + h.c. \right\}.
\end{eqnarray}
The distance $|z_1-z_2|$ between the superposed
coherent states determines the strength of the perturbation of the echo
Hamiltonian (\ref{echoH}). Thus, fidelity
decay (and decoherence) become more rapid, as this distance increases.
Assuming Markovian behavior and $\gamma t\ll 1$, our result reduces to the
famous relation $|\varrho_{12}^{\rm c}(t)|^2
= e^{-\gamma t|z_1(0)-z_2(0)|^2}\; |\varrho_{12}^{\rm c}(0)|^2$
\cite{Zeh,Zurek,ZurHab93}.

\section{\label{S} Situations where product state solutions are
only approximate}

As the previous examples have shown, the relation
between decoherence in the central system and fidelity
decay in the environment works nicely, whenever it is
possible to find product-state solutions of the coupled dynamics.
In general, this will not be possible. Approximate
product state solutions lead to approximate pure state
solutions of the reduced dynamics and therefore to the
concept of ``robust'' or ``pointer'' states \cite{Zeh,Zurek,ZurHab93}.
Thus, if pointer states may be identified, the decoherence-fidelity
relation will be satisfied in an approximate sense. A
detailed discussion is beyond the scope of the
current paper. The following short-time analysis of decoherence,
however, allows for the desired relation in a very common situation.

\subsection*{Short time approach to decoherence}

As the ``distance'' between superposed quantum states grows,
decoherence may become very rapid. This observation is the starting point
of a general short-time approach to decoherence recently developed
\cite{Bra00,SHB03,SH03}. Let us briefly sketch the main ideas.

Consider a central system with Hamiltonian $H^{\rm c}$
coupled to the environment (Hamiltonian $H^{\rm e}$)
through an interaction of the form
\begin{equation}
H^{\rm int} = S\otimes V
\end{equation}
where $S$ ($V$) is some operator in the Hilbert space of the central
system (of the environment). Typically, the environmental
part consists of contributions of many independent degrees of freedom,
$V=\sum_\lambda V_\lambda$, but this is not of importance here.
Decoherence in the central system will be most effective for
initial states with largely different expectation values of $S$.
In the famous quantum Brownian motion case \cite{Zeh,Zurek, CalLeg},
for instance,
we have $H^{\rm c} = \frac{p^2}{2m} + V^{\rm c}(q)$,
and $S=q$, the position operator.

In analogy to the oscillator case we assume an initial state
of the form
\begin{equation}\label{sinitial}
|\Psi(0)\rangle = \frac{1}{\sqrt{2}}(|s\rangle + |s'\rangle)\otimes
|B(0)\rangle \; ,
\end{equation}
where $|s\ra , |s'\ra$ and $s,s'$ are eigenstates and corresponding eigenvalues
of the operator $S$.
In \cite{Bra00,SHB03,SH03} it is argued that if $|s-s'|$ is large
enough, decoherence may be so rapid as to outrun any
dynamics induced by the Hamiltonian $H^{\rm c}$ of the central system.
Thus, for these short times, $H^{\rm c}$ may be dropped entirely and
the total Hamiltonian reads
\begin{equation}
H \approx   S\otimes V + H^{\rm e}.
\end{equation}
In this short-time approximation, eigenstates of $S$ are conserved.
Thus, we are essentially in the ``dephasing'' situation, discussed in
section~\ref{E}. Again, one finds product state solutions of
the Schr\"odinger equation, here of the form
\begin{equation}
|\Psi(t)\rangle = |s\rangle\otimes|B_s(t)\rangle.
\end{equation}
The environmental evolution is generated by the Hamiltonian
\begin{equation}
H^{\rm e}_s =  H^{\rm e} + sV.
\end{equation}
For the environmental dynamics,
the eigenvalue $s$ plays the role of a coupling strength
to the ``potential'' $V$.
 Thus, for times shorter than any time scale induced by the central
Hamiltonian $H^{\rm c}$, the solution of the Schr\"odinger equation
 for the initial state (\ref{sinitial}) will be
 \begin{equation}\label{sentang}
 |\Psi(t)\rangle = \frac{1}{\sqrt{2}} |s\rangle\otimes|B_s(t)\rangle
 + \frac{1}{\sqrt{2}} |s'\rangle\otimes|B_{s'}(t)\rangle
 \end{equation}

The reduced density operator of the central
system follows similarly to expression (\ref{reddens})
and we conclude that the coherence between the states
$|s\rangle$ and $|s'\rangle$ is given by
the overlap
\begin{eqnarray}\label{shorttime}
\varrho^{\rm c}_{s\,s'}(t) & = &
\langle s|\varrho^{\rm c}(t)|s'\rangle =
\langle B_{s'}(t)| B_{s}(t)\rangle
\;\varrho^{\rm c}_{s\,s'}(0)\\ \nonumber
& = & \langle B(0)|{\tilde U}_0^\dagger(t){\tilde U}(t)| B(0)\rangle
\;\varrho^{\rm c}_{s\,s'}(0). \nonumber
\end{eqnarray}
Here, the propagators correspond to the environment Hamiltonians
\begin{equation}
{\tilde H}_0 = H^{\rm e} + s'V\;\;\;\;\mbox{and}\;\;\; {\tilde H} =
{\tilde H}_0 + (s-s')V.
\end{equation}
Decoherence in $\varrho^{\rm c}$
may thus be interpreted as an echo fidelity with a perturbation
proportional to the difference of the eigenvalues $s-s'$ of the
initially superposed states. We recall and stress that by self
consistency, this simple short-time result is valid only as long as
it predicts decoherence (fidelity decay) times that are short compared
to ``system'' time scales induced by the central Hamiltonian.

\section{\label{C} Conclusions and Outlook}

We have analyzed the connection between decoherence of a central system and
fidelity decay in the environment for a variety of situations.
This connection can be established easily if the energy of the central system
is conserved ({\it i.e.} dephasing) as also discussed in~\cite{GCZ97,GCZ97b}.
Here, we have extended these ideas to more general
situations. Interestingly, we have been able to show that even in the case of
dissipation (amplitude coupling) a similar relation holds. Moreover,
short time decoherence can be interpreted along these lines.
Generally speaking, for our argument to be valid it is crucial that the
Hamiltonian of the composite system allows for (approximate) product state
solutions as time evolves. Then, the superposition of two such solutions allows
to interpret the decoherence manifest in the off-diagonal
matrix element of the reduced density matrix of the central system as a
fidelity decay in the environment and vice versa.
It is remarkable that properties of unitary time evolution in the environment
and the non-unitary evolution of coherences become related.

Experiments based on these ideas can give important information
about the stability of the unperturbed (environmental) Hamiltonian -- a fact
which might also be relevant for quantum information processing.

The connection between decoherence and fidelity decay can always be established
whenever pointer states of the central system can be found. Then the
factorization, essential to our argument, is valid for fairly long times.
%
%
Our results highlight a beautiful complementarity: decoherence between pointer 
states may be interpreted as an act of measurement by the environment on
the central system. The ``collapsed'' state of the central system may then be
inferred from the environment. When measuring fidelity decay via decoherence,
information about the environmental dynamics is extracted via observations on
the central system. Implications of this connection will have to be studied 
in future work.

\section*{Acknowledgments}

Financial support by 
CONACyT project \#41000F and DGAPA-UNAM project \#IN100803
is acknowledged. Two of us (T.G. and T.P.) are grateful
for the hospitality at the Centro Internacional de Ciencias,
Cuernavaca, Mexico. T.P. is financially supported by the grants
P1-0044, Ministry of Science, Education and Sport, Slovenia, and
DAAD19-02-1-0086, ARO United States. T. G. acknowledges financial support
by the EU Human Potential Program, contract HPRN-CT-2000-00156.

\end{document}